\documentclass[12pt]{iopart}

\usepackage{bm,bbm}
\usepackage{graphicx}
\usepackage{amsmath}
\usepackage{amssymb}
\usepackage[T1]{fontenc}

\newcommand{\vb}[1]{\bm{ #1}}

\newcommand{\ie}{i.e. }

\newcommand{\unite}[1]{\,\mathrm{#1}}
\newcommand{\dd}{{\textstyle \frac{1}{2}}}
\newcommand{\der}{{\textrm{d}}}
\newcommand{\op}[1]{\mathrm{#1}}
\newcommand{\T}{\mathcal{T}}
\newcommand{\cchi}{\mathbbm{1}}

\begin{document}

\title[Background estimation by time slides]{On the background
  estimation by time slides in a network of gravitational wave
  detectors}

\author{Micha\l{} W\k{a}s, Marie-Anne Bizouard, Violette Brisson, Fabien 
Cavalier, Michel Davier, Patrice Hello, Nicolas Leroy, Florent 
Robinet and Miltiadis Vavoulidis.}

\address{LAL, Univ Paris-Sud, CNRS/IN2P3, Orsay, France.}
\ead{mwas@lal.in2p3.fr}
\begin{abstract}
  Time shifting the outputs of Gravitational Wave detectors operating
  in coincidence is a convenient way to estimate the background in a
  search for short duration signals.  However this procedure is
  limited as increasing indefinitely the number of time shifts does
  not provide better estimates. We show that the false alarm rate
  estimation error saturates with the number of time shifts. In
  particular, for detectors with very different trigger rates this
  error saturates at a large value. Explicit computations are done 
  for 2 detectors, and for 3 detectors where the detection statistic
  relies on the logical ``OR'' of the coincidences of the 3 couples in
  the network.
\end{abstract}

\maketitle

\section{Introduction}

Kilometric interferometric Gravitational Wave (GW) detectors such as
LIGO \cite{LIGO} or Virgo \cite{Virgo} have been taking data with
increasing sensitivities over the past years
\cite{burstS4,burstS5,burstS5HF,cbcS4,lowMassS5,lowMassS5y2,virgoC7}.  It is expected that short
duration GW events, e.g. the so-called bursts emitted by gravitational
collapses or the signals emitted by compact binary inspirals, are very
rare.  Moreover the output of the detectors is primarily (non
Gaussian) noise, and this background noise is in general not
modeled. This implies that with a single GW detector it is very
difficult to estimate the background event rate, and then to assess the
significance of some GW candidate.

On the contrary when dealing with a network of detectors (that means
in practice at least two detectors of the same class), there is a
conventional and simple way to estimate the background, that is the
rate of coincident events due to detector noise. This consists
of time shifting the search algorithm outputs (or triggers) of each
detector with respect to the other(s), by some unphysical delays, much
larger than the light travel time between the detectors and much
larger than the typical duration of an expected GW signal. The next
step is then to look for coincidences between shifted triggers just as
if the shifted streams were synchronized.
% All the analysis cuts are then defined with these time shifted
% ``events'', \ie on the background.
As we deal with \emph{a priori} rare events, we need to set in
practice low false alarm rates in the analysis. The question then
arises of how many time slides are needed for correctly estimating the
background and especially its tails where rare (non-Gaussian noise) events lay.
% and then necessary to estimate the significance of some event against the background.
Note, that in practice in burst or binary inspiral searches, a
hundred or more time slides are done \cite{burstS4,cbcS4,lowMassS5},
due in particular to limited computational resources. Such a
limitation of course depends on the duration of the different
detectors data streams and on the complexity of consistency tests
performed on coincident triggers.  For example, time slides
computation for one year of data sampled at 16384 Hz (LIGO) or 20 kHz
(Virgo) can rapidly become a computational burden, especially when
computationally intensive consistency tests like the $\chi^2$ veto \cite{Allen05}
are used.

In this paper, we show that the precision on the background
estimation, using time slides of trigger streams, is in fact
limited. Indeed the variance of the false alarm rate estimation does
not indefinitely decrease as the number of time slides increases as we
would naively believe. On the contrary this variance saturates at some
point, depending on the trigger rates chosen in each individual
detector and on the coincidence time window set for identifying
coincident events in the network of detectors.  

After introducing the general definitions in section
\ref{sec:definitions}, we give explicit formulas for the
two-detector and the three-detector case in respectively
Sec.~\ref{sec:2det} and \ref{sec:3det}. In the latter we restrict
ourselves to the particular analysis scheme where we are looking at
the union (logical ``OR'') of the three couples of detectors. This is
actually the configuration of interest, since it is more sensitive
than simply searching for triple coincidences (logical ``AND'')
\cite{p1b}. In each case (2 or 3 detectors) we check the analytical
result with a Monte Carlo simulation and find excellent agreement. In
section \ref{sec:Discussion} these results are applied and discussed
using typical parameters of a GW data analysis.

\section{Definitions}\label{sec:definitions}
\subsection{Poisson approximation for trigger generation}
Background triggers in the detectors are due to rare glitches. Often
these glitches come in groups, but most analysis pipelines cluster
their triggers, so each glitch group results in only one final
trigger. This clustering procedure is reasonable as long as the
resulting trigger rate is much lower than the inverse of the typical
clustering time length. In this limit the clustered
triggers are independent events. Thus, throughout this paper we will
assume that each detector produces random background triggers, which
are Poisson distributed in time.

\subsection{Problem description}
We look then at the coincidence between two Poisson
processes. The single interferometer trigger rate will be noted
$\op{FA}_1$, $\op{FA}_2$, ..., the coincidence rate will be simply noted
$\op{FA}$. We denote by $\widetilde{\op{FA}}(\T)$ the rate resulting from counting
the number of coincidence between two data streams, that are shifted
by some time $\T$.  In particular for zero lag ($\T=0$) the measured rate is
$\widetilde{\op{FA}}(0)$. So the quantities with tildes are the
experimentally measured rates, and the quantities without tildes are
the actual Poisson distribution parameters. The purpose of the paper
is to study the properties of the time shifting method, which uses
$\widehat{\op{FA}}=\frac{1}{R}\sum_{k=1}^R \widetilde{\op{FA}}(\T_k)$ as an estimator
of $\op{FA}$, where $R$ is the number of time slides and $\T_k$ is the
$\text{k}^{\text{th}}$ time slide stride. 

\subsection{Poisson process model}
To model a Poisson process with event rate $\op{FA}_1$ we discretize
the data stream duration $T$ with bins of length $\Delta t$, the
discretization time scale, e.g. either the detector sampling rate or the
clustering time scale.  Thus,
for each bin an event is present with a probability $p = \op{FA}_1
\Delta t$ \footnote{Here we model the Poisson process by a binomial
  distribution, recalling that when $p\ll1$ the binomial distribution tends
  toward a Poisson distribution}.  %The time $T$ corresponds to the
%data stream length and $\Delta t$ is the discretization time scale,
%e.g. the detector sampling rate or the clustering time scale.
%For two Poissonian processes with event rate $\op{FA}_1$ and $\op{FA}_2$, we
%have a coincidence if there is an event in the same time bin $k$ for
%both processes.

To ease the calculation we describe the Poisson process
realizations with a continuous random variable. We take $\vb{x}$
uniformly distributed in the volume $[0,1]^{N}$ where
$N=\frac{T}{\Delta t}$ is the number of samples, then compare $x_k$
(the $\text{k}^{\text{th}}$ coordinate of $\vb{x}$) with $p$. When
$x_k<p$ there is an event in time bin $k$, otherwise there is
none. Thus $\vb{x}$ characterizes one realization of a Poisson
process, and it can be easily seen that the uniform distribution of
$\vb{x}$ leads to a Poisson distribution of events.

\subsection{Coincidences}\label{sec:definitions:coincidences}
We choose the sampling $\Delta t$ to be equal to twice the coincidence
time window $\tau_c$, in order to simplify the modeling of the
coincidence between two processes. More precisely, for two Poisson
processes with event rates respectively $\op{FA}_1$ and $\op{FA}_2$,
we define a coincidence when there is an event in the same time bin $k$
for both processes. This is different from the usual definition, where
events are said in coincidence when they are less than a time window
apart. This binning time coincidence has on average the same effects
as defining as coincident events that are less than $\pm\dd
\Delta t = \pm \tau_c$ apart. The analytical results are derived using
this non standard definition, but they are in precise agreement with
Monte Carlo simulations that are performed using the usual definition
of time coincidence.

%   As we will confirm with the Monte Carlo
% simulations, the analytical results derived with this modified
% definition remain exact when applying them to the usual definition of
% the time coincidence.

\section{The case of two detectors}
\label{sec:2det}
\subsection{Time slides between two detectors}\label{sec:two detector case}

Let $\vb{x}, \vb{y} \in [0,1]^N$ be two realizations of Poisson processes with
respectively $p = \op{FA}_1 \Delta t$ and $q = \op{FA}_2 \Delta t$.
There is a coincident event in time bin $k$ when $x_k < p$ and $y_k <
q$. So the total number of coincidences for this realization is
\begin{equation}
 \sum_{k=1}^N \cchi(x_k<p) \cchi(y_k<q)
\end{equation}
where 
\begin{equation}
 \left\{
\begin{matrix}
  \cchi(a) = 1 &\qquad \text{if } a \text{ is true} \\
  \cchi(a) = 0 &\qquad \text{if } a \text{ is false}
 \end{matrix}
\right.
\end{equation}
Thus the mean number of coincidences without time slides is as expected
\begin{equation}
 \mathrm{Mean} = \int_{x_1} \dotsi \int_{x_N} \int_{y_1} \dotsi \int_{y_N} \sum_{k=1}^N \cchi(x_k<p) \cchi(y_k<q) 
\underbrace{\der x_1 \dotsm \der x_N \der y_1 \dotsm \der y_N}_{\der V}= N p q.
\end{equation}

To consider a number $R$ of time slides we take a set of $R$ circular
permutations of $\lshad1,N\rshad$. Time-sliding a vector $\vb{x}$ by
the circular permutation $\pi$ transforms the vector $\vb{x}$ into the
vector of coordinates $x_{\pi(k)}$. Then the mean number of
coincidences is simply
\begin{equation}
 \mathrm{Mean} = \int_{x_1} \dotsi \int_{x_N} \int_{y_1} \dotsi
 \int_{y_N} \frac{1}{R} \sum_\pi \sum_k \cchi(x_k<p) \cchi(y_{\pi(k)}<q)
 \der V = N p q,
\end{equation}
thus there is no bias resulting from the use of time slides.

\subsection{Computation of the variance}
\begin{figure}
\begin{center}
\includegraphics[width=0.8\textwidth]{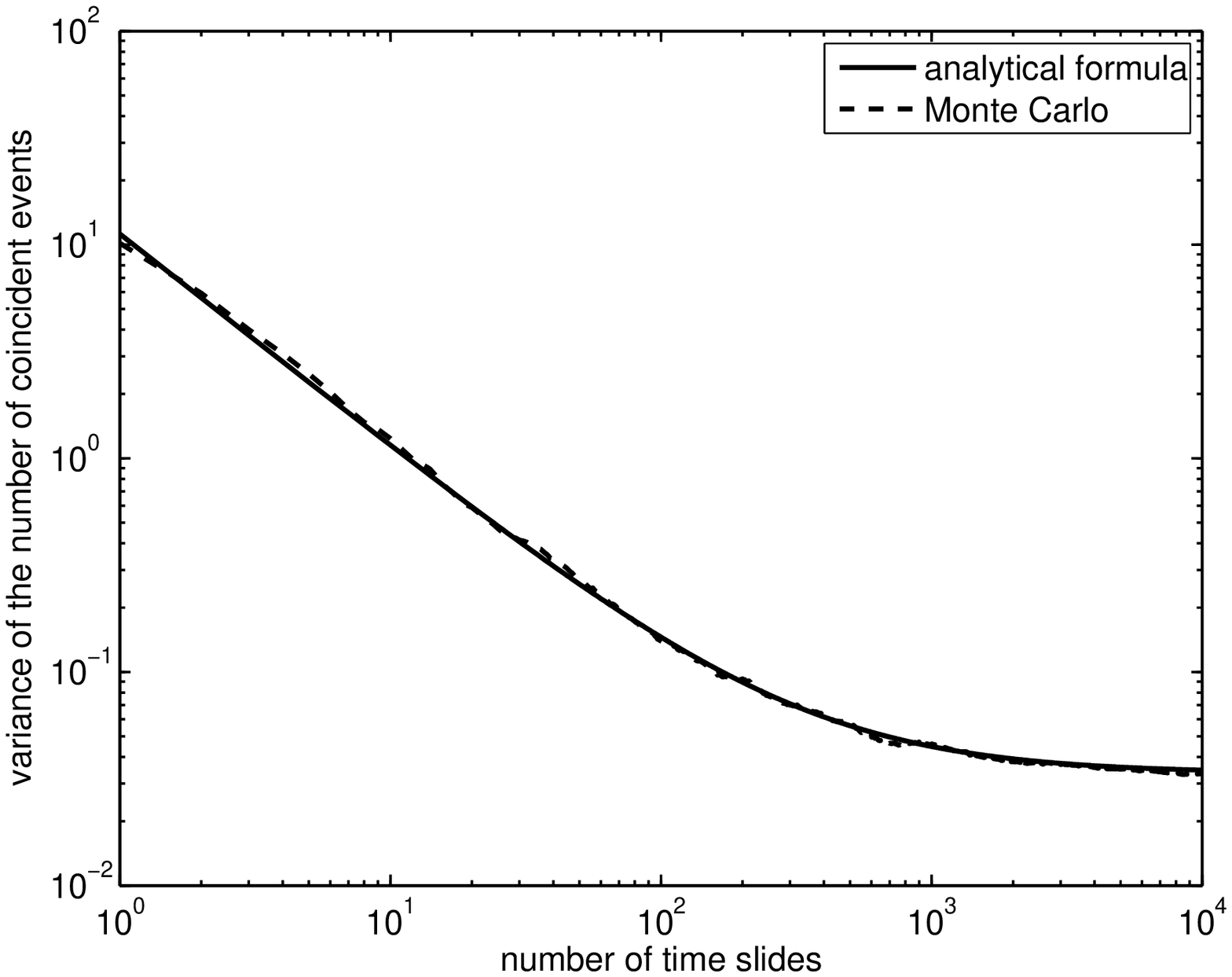}
\end{center}
\caption{\label{thVsExp} The solid line is the analytical formula
  \eqref{varNb} of the variance and the dashed line is the Monte Carlo
  variance as a function of the number of times slides. The Monte
  Carlo has been performed with $\op{FA}_1 = 0.7\unite{Hz},\,
  \op{FA}_2 = 0.8\unite{Hz},\, \tau_c = 1\unite{ms}$, 500 trials and a
  $T=10^4\unite{s}$ data stream length. }
\end{figure}

In order to have an estimate of the statistical error, we compute the
variance with $R$ time slides. The second moment is
\begin{multline}
 \op{M}_2  = \int_{x_1} \dotsi \int_{x_N} \int_{y_1} \dotsi \int_{y_N}
 \left[ \frac{1}{R}\sum_\pi \sum_k\cchi(x_k<p) \cchi(y_{\pi(k)}<q)
 \right]^2 \der  V \\
  = \int \dotsi \int \frac{1}{R^2} \sum_{\pi_1} \sum_{\pi_2} \sum_{k} \sum_{l} 
\cchi(x_{k}<p) \cchi(x_{l}<p) \cchi(y_{\pi_1(k)}<q) \cchi(y_{\pi_2(l)}<q)
\der V .
\end{multline}
We can then exchange integrals and sums. To compute the integrals we
distinguish two cases: when $k\neq l$ the integrals on $x_k$ and $x_l$
are independent, and the integration over $x_1, \dotsc, x_N $ gives a
$p^2$ contribution; otherwise the integration gives a $p$
contribution. Analogously for the $y$ variables we obtain $q$ or $q^2$
depending on whether $\pi_1(k) = \pi_2(l)$ or not.

The computation of this integral, detailed in \ref{twoITFcalc}, yields
\begin{align}
  \mathrm{Var }
  & = N p q \left[ \frac{1}{R} + p + q + \frac{p q - (p + q)}{R} - 2 p q\right] \\
  & \simeq N p q \left[ \frac{1}{R} + p + q\right] \label{varNb},
\end{align}
where the last line is an approximation in the limit
$p,\,q,\,\frac{1}{R} \ll1$, which is reasonable as far as GW analysis
is concerned.

\subsection{Interpretation}\label{sec:2det:interpretation}
 
Each term in equation \eqref{varNb} can be interpreted. The
$\frac{1}{R}$ is what we would expect if we considered $R$ independent
Poisson process realizations instead of $R$ time slides. The $p+q$
comes from the estimation of the Poisson process event rate. Indeed,
the estimation of the event probability $p$ from a single realization
of a Poisson process with a mean number of events $N p$ is
$\widehat{p} = p + \delta p$, where $\delta p$ is the random
statistical error with variance $\langle \delta p^2\rangle =
\frac{p}{N}$. This yields the mean rate of coincidences
\begin{equation}\label{eq:interpretation}
  \text{Mean} = N \widehat{p} \widehat{q} = N \left(p + \delta p \right)\left(q  + \delta q\right
  ) \simeq N p q + N p \delta q + N q \delta p
\end{equation}
which corresponds to a variance of $\left\langle N^2 p^2 \delta q^2 + N q^2 \delta
p^2 \right\rangle= N p q (p + q)$, because $\delta p$ and $\delta q$ are
independent errors. Thus, when using only one realization for the
single detector triggers, we have a statistical error on the single
detector process rate. This statistical error is systematically
propagated to the coincidence rate of each time slide, that yields the
extra terms in the variance as compared to independent process
realizations. One can see that this extra term is important when
$\frac{1}{R} < \max(p,q)$; for cases where the coincident false alarm
rate is maintained fixed ($pq$ constant), the effect is most
noticeable when $p$ and $q$ are very different.

This gives an estimate of the variance of the number of coincident
events in a data stream of length $T$. After converting to the
estimation of the coincidence false alarm rate we obtain
\begin{align}
  \mathrm{Mean}_{\widehat{\op{FA}}} & = \frac{\mathrm{Mean}}{T} = \op{FA}_1 \op{FA}_2 \Delta t, \\
  \mathrm{Var}_{\widehat{\op{FA}}} & = \frac{\mathrm{Var}}{T^2} \simeq \op{FA}_1 \op{FA}_2 \frac{\Delta t }{T} \left[ \frac{1}{R} + \op{FA}_1 \Delta t + \op{FA}_2 \Delta t\right]. \label{varFA2}
\end{align}

%\subsubsection{Monte Carlo verification}\label{2detector:MC}

To verify these results, a Monte Carlo simulation has been
performed. The Poisson processes are created as described in section
\ref{sec:definitions}, using a sampling rate of $16384 \unite{Hz}$,
then a simple coincidence test with a window of $\tau_c = 1
\unite{ms}$ is applied. The time shifts are done by adding an integer
number of seconds to all events and applying a modulo $T$ operation.
The formula has been tested using 500 realizations of $T = 10^4
\unite{s}$ long Poisson processes, and using between 1 and $10^4$ time
slides for each realization. Figure \ref{thVsExp} shows that the
analytical formula \eqref{varNb} and the Monte Carlo agree well for
any number of time slides, and that the variance starts saturating
when a few hundred time slides are used. We can see that the
identification of the sampling time and the coincidence time window
has no consequence on the result, the choice between binning and
windowing the coincidences is a higher order effect.

\subsection{Straightforward extensions of the model}
%For the moment we assumed that the trigger streams have no holes. 
In real data analysis, there are times when one of the detectors does not
take science quality data for technical reasons. Thus, the data set is
divided into disjoint segments, and the background estimation is often
done by circular time slides on each segment separately. Afterwards
the results from all the segments are combined to get the background
false alarm estimation. The computation discussed above extends to this
case with minimal changes. The circular permutations have to be changed
to circular by block permutations, everything else being kept
identical.

Another caveat is that for real data analysis the coincidence
procedure is often more complicated. Some of those complications are
event consistency tests, e.g. do the two coincident events have a similar
frequency? We can model this by adding some parameter $f$ distributed
uniformly in $[0,1]$ attached to each event, and then requesting a
coincidence in the parameter $f$.

For this model the results will be the same as those above, up to a
factor of order 1. Indeed, instead of applying a window of size
$\Delta t$ to our events, we are now working in a 2 dimensional (for
instance time-frequency) space and using a rectangular window in this
2D parameter space. The procedure in both cases is the same ---
applying D dimensional rectangular windows to events distributed
uniformly in a D dimensional space --- up to the dimension of the
space.

\section{The case of three detectors}
\label{sec:3det}
\subsection{Time slides between three detectors}
In the case of three detectors one natural extension is to ask for
events that are seen by at least two detectors, which means look for
coincidence between two detectors for each detector pair, but counting
the coincidences between three detectors only once. This ``OR''
strategy in a interferometer network has been shown to be more
efficient than a direct three fold coincidence strategy (``AND''
strategy) \cite{p1b}. For time slides, when shifting the events of the
second detector with some permutation $\pi$, we also shift the events of the third
detector by the same amount but in the opposite direction with
$\pi^{-1}$. To write compact equations we abbreviate
$X=\cchi(x_k<p)$, $Y=\cchi(y_{\pi(k)}<q)$, $Z=\cchi(z_{\pi^{-1}(k)}<r)$,
$\der V = \der x_1 \dotsm \der x_N \der y_1 \dotsm \der y_N \der z_1
\dotsm \der z_N $, where $r=\op{FA}_3 \Delta t$ is the event
probability per bin of the third detector and the vector $\vb{z}$
describes its realizations. Thus, the mean number of coincidences in the
framework described in section \ref{sec:two detector case} is
%\begin{multline}
% \mathrm{Mean} = \int_{x_1} \dotsi \int_{x_N} \int_{y_1} \dotsi \int_{y_N} \int_{z_1} \dotsi \int_{z_N} \frac{1}{R} \sum_\pi \sum_k \der x_1 \dotsm \der x_N \der y_1 \dotsm \der y_N \der z_1 \dotsm \der z_N \\ \left[\cchi(x_k<p) \cchi(y_{\pi(k)}<q)  +  \cchi(x_k<p) \cchi(z_{\pi^{-1}(k)}<r) +   \cchi(y_{\pi(k)}<q) \cchi(z_{\pi^{-1}(k)}<r) \right.  \\ \left. - 2 \cchi(x_k<p) \cchi(y_{\pi(k)}<q) \cchi(z_{\pi^{-1}(k)}<r)\right] 
%\end{multline}
%or more compactly
\begin{equation}
 \op{Mean}= \int \dotsi \int \frac{1}{R} \sum_\pi \sum_k \left[ X Y +
   Y Z + X Z - 2 X Y Z\right] \der V= N \left[ p q + p r + q r - 2 p q
   r\right].
\end{equation}

\subsection{Computation of the variance}

The second moment can be written compactly as
\begin{multline}
 \op{M}_2 = \int \dotsi \int \frac{1}{R^2} \sum_{\pi_1} \sum_{\pi_2} \sum_k \sum_l 
\\\left[ X Y X' Y' + X Z X' Z' + Y Z Y' Z' + 4XYZX'Y'Z' + 2XYX'Z' +
  2XYY'Z' \right. \\ \left. + 2XZY'Z' -4 XYX'Y'Z' - 4XZX'Y'Z' -
  4YZX'Y'Z'\right] \der V,
\end{multline}
where the $'$ denotes whether the hidden variables are $\pi_1,\,k$ or $\pi_2,\,l$. 

The computation of this integral, detailed in \ref{threeITFcalc}, yields
\begin{multline}
 \op{M}_2 =\frac{N}{R}\Bigl\{ (pq + pr + qr- 2pqr) \Bigr. \\ \left.+ (R-1)\left[ pq(p+q+pq) + pr(p+r+pr) +qr(q+r+qr) + 6pqr -4pqr(p+q+r)\right]  \right. \\ \Bigl.+ \left[ (R-1)(N-3)+(N-1) \right](pq+pr+qr-2pqr)^2\Bigr\},
\end{multline}
and can be approximated in the limit $p,q,r,\frac{1}{R} \ll 1$ by 
\begin{equation}\label{varNb3}
 \op{Var} \simeq N(pq + pr + qr)\left( \frac{1}{R} + p + q + r +
   \frac{3 p q r}{pq + pr + qr} \right).
\end{equation}

\subsection{Interpretation}\label{sec:3det:interpretation}
\begin{figure}
\begin{center}
\includegraphics[width=0.8\textwidth]{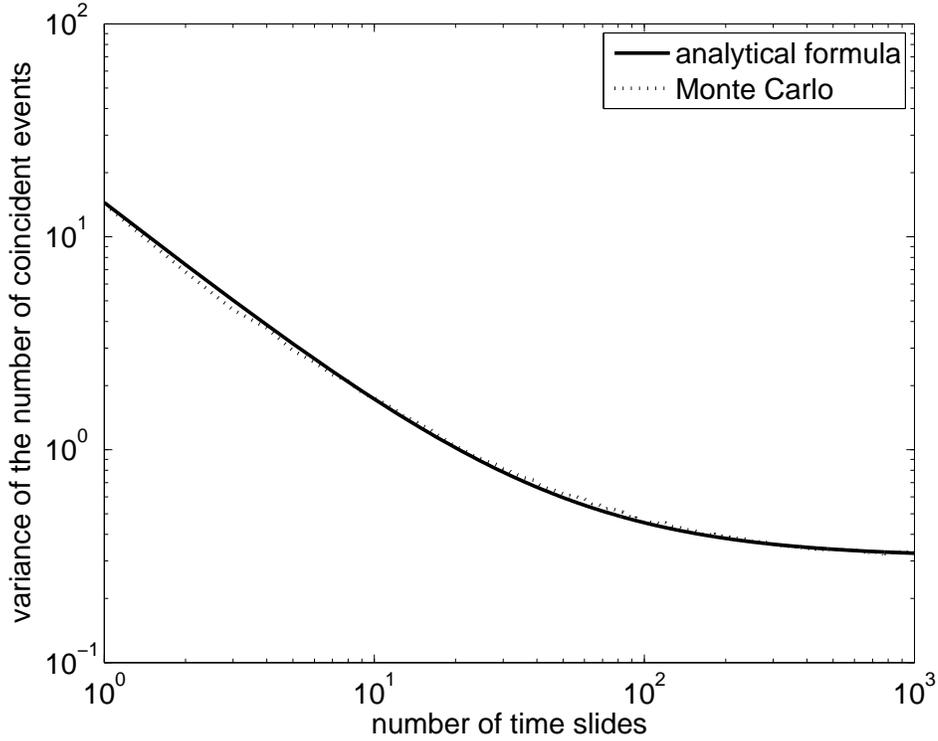}
\end{center}
\caption{\label{thVsExp3ITF} The solid line is the analytical formula
  \eqref{varNb3} of the variance and the dotted line is the Monte
  Carlo variance as a function of the number of time slides. The Monte
  Carlo has been performed with $\op{FA}_1 = 0.04\unite{Hz},\,
  \op{FA}_2 = 0.08\unite{Hz},\, \op{FA}_3 = 0.16\unite{Hz},\, \tau_c =
  31\unite{ms}$, 500 trials and a $T=10^4\unite{s}$ data stream
  length.}
\end{figure}

Similarly to section \ref{sec:2det:interpretation} the extra terms in
equation \eqref{varNb3} can be explained through the error in the
estimation of the single detector event rate. Using the same notations
as in section \ref{sec:2det:interpretation} the mean coincidence number is
\begin{equation}
  \label{eq:3det:interpretation}
  \text{Mean} = N \left( \hat{p} \hat{q} + \hat{p} \hat{r} + \hat{q}
    \hat{r} \right) \simeq
  N \left[ p q + p r + q r  + (q + r)\delta p + (p + r)\delta q + (p +
  q) \delta r\right].
\end{equation}
Using the independence of estimation errors and recalling
that $\langle \delta p^2\rangle = \frac{p}{N}$ we obtain the variance
of this mean value
\begin{align}
  \text{Var} &= N^2 \left[\langle \delta p^2\rangle (q+r)^2 + \langle
    \delta q^2 \rangle (p + r)^2 + \langle \delta
    r^2\rangle  ( p + q)^2 \right]  \notag\\
  &= N \left[ ( p q + p r + q r) (p + q + r ) + 3 p q r \right],
\end{align}
that corresponds to the extra terms in equation \eqref{varNb3}.

After converting to the estimation of the false alarm rate we obtain
\begin{align}
 \mathrm{Mean}_{\widehat{\op{FA}}}  & \simeq (\op{FA}_1 \op{FA}_2 + \op{FA}_1 \op{FA}_3+ \op{FA}_2 \op{FA}_3)\Delta t, \\
 \mathrm{Var}_{\widehat{\op{FA}}} & \simeq (\op{FA}_1 \op{FA}_2 + \op{FA}_1 \op{FA}_3+ \op{FA}_2 \op{FA}_3) \frac{\Delta t }{T} \notag \\
& \phantom{\simeq} \left( \frac{1}{R} + \op{FA}_1 \Delta t + \op{FA}_2 \Delta t  + \op{FA}_3 \Delta t  + 
\frac{3 \op{FA}_1 \op{FA}_2 \op{FA}_3 }{\op{FA}_1 \op{FA}_2 +
  \op{FA}_1 \op{FA}_3+ \op{FA}_2 \op{FA}_3} \Delta t\right).
\end{align}

%\subsubsection{Monte Carlo verification}

To check the 3 detector results we performed a Monte Carlo similar to
the one of the 2 detector case (see section \ref{sec:2det:interpretation}). The
only difference is the number of detectors, and we choose a different
coincidence window: $\tau_c =31 \unite{ms}$ \footnote{This
  accounts for the largest light travel time in the LIGO-Virgo network
  ($27\unite{ms}$) and some timing error in each detector.}.  To check
that the assumption of equal and opposite time slides does not
influence the result, in the Monte Carlo the data in the second
detector are shifted by $\T_k$ and in the third detector by $3 \T_k$. Figure
\ref{thVsExp3ITF} shows that the Monte Carlo and the 3 detector ``OR''
formula \eqref{varNb3} agree really well.

\subsection{The case of $D$ detectors}
For the sake of completeness we can generalize the interpretation done
in section \ref{sec:2det:interpretation} to the case of $D$ detectors in
the ``AND'' configuration. This generalization of equation
\eqref{eq:interpretation} to $D$ detectors yields a variance on the
number of coincidences
\begin{equation} \label{varDdet}
  \op{Var} \simeq N \prod_{i=1}^D p_i \left(\frac{1}{R} + \sum_{i=1}^D \prod_{j \neq i} p_j \right),
\end{equation}
where $p_i$ is the probability for detector $i$ to have an event in
a given time bin.
% We checked this result for the three detectors case with a detailed
% computation similar to that of \ref{threeITFcalc}. For higher $D$ the
% detailed computation becomes quickly cumbersome, but we expect this
% interpretation to stay true for any $D$.

The interpretation can also be generalized in the ``OR'' case, that is
coincidence between any pair of detectors, although the computation is
more cumbersome as detailed in \ref{sec:orDdim} and yields
\begin{equation}
  \label{eq:orDdim}
  \op{Var} \simeq N \left[ \left( \sum_{i<j} p_i p_j \right) 
    \left( \frac{1}{R} + \sum_{i=1}^D p_i  \right)
    + \frac{1}{2} \sum_{i \neq j,\, j \neq k,\, k \neq i} p_i p_j p_k
\right], 
\end{equation}
where $p_i$ is the event probability per bin in the
$\text{i}^\text{th}$ detector.

\section{Discussion}\label{sec:Discussion}

We finally discuss the consequences of the above results on GW data
analysis. To be able to put numbers into the equations we will look at
a fiducial GW data taking run.  We choose the run properties to be:
\begin{itemize}
\item {a duration of $T=10^7\unite{s}$, that is roughly 4 months}
\item {two detectors with a light travel time separation of
    $25\unite{ms}$, and we use the same time as the coincidence
    window, so that $\Delta t = 50 \unite{ms}$\footnote{As noted in
      section \ref{sec:definitions:coincidences}, coincident triggers
      are defined as less that $\pm \dd \Delta t$ apart.}, assuming perfect
    timing accuracy of trigger generators.}
\item {a desired coincidence false alarm rate of $10^{-8}\unite{Hz}$,
    \ie one event every three years}.
\end{itemize}

We will look at two special cases of single detector threshold
choice. One symmetric case, where thresholds are set so that the
single detector trigger rate in each detector is roughly the same. One
asymmetric case, where in one of the detectors there is only one
trigger. This asymmetric case is extreme but instructive, because
tuning the thresholds to obtain the best sensitivity often yields
asymmetric trigger rates between different detectors.

\begin{description}
\item[Symmetric detector case] In this case we have the single
  detector trigger rate $ \op{FA}_1 = \op{FA}_2 = \op{FA}_s =
  \sqrt{\frac{\op{FA}}{\Delta t}} \simeq 4.5 \times
  10^{-4}\unite{Hz}$, which gives using equation \eqref{varFA2} the
  fractional error of the false alarm estimation
\begin{align*}
 \frac{\sigma_{\op{FA}}}{\op{FA}} &\simeq 3.2 \left(\frac{1}{R}+ 4.5
   \times 10^{-5} \right)^{\dd} & & p = q = 2.25 \times 10^{-5}\\
&\simeq 0.32 & & \text{for } R=100 \\
&\simeq 0.02 & & \text{for } R \rightarrow \infty.
\end{align*}
So for 100 time slides we get a typical error of 30\% in the false alarm  estimation, and  the error saturates at 2\% for $R \gtrsim 20000$.

\item[Asymmetric detector case] In this extreme case the single
  detector trigger rates are $\op{FA}_1=\frac{1}{T}=10^{-7}\unite{Hz}$
  and $\op{FA}_2=\frac{\op{FA}}{\op{FA}_1 \Delta t} = 2\unite{Hz}$,
  which gives using equation \eqref{varFA2} the fractional error of
  the false alarm estimation
\begin{align*}
 \frac{\sigma_{\op{FA}}}{\op{FA}} &\simeq 3.2 \left(\frac{1}{R}+
   0.1\right)^{\dd} & &p=5\times 10^{-9},\; q=0.1\\
&\simeq 1.05 & &\text{for } R=100 \\
&\simeq 1 & &\text{for } R \rightarrow \infty.
\end{align*}
So the error saturates at 100\%, and this saturation is achieved for $R \gtrsim 10$.
\end{description}

Those two examples show that the maximal number of useful time slides
and the false alarm estimation precision strongly depends on the
relative properties of the two detectors. In particular when there are
much more triggers in one detector than in the other, the background
can be badly estimated and increasing the number of time slides does
not solve the issue.

%Let us note that the time slide method and the results presented above are not specific to GW data analysis. They can be used for any experiment where coincidences between two (or more) trigger generator are looked for and the results
%of this paper can be straightforwardly extended to such an experiment. 

\section{Conclusions}

We have studied the statistical error in the background
estimation of event-based GW data analysis when using the time slide
method. Under the assumption of stationary noise we analytically
computed this error in both the two-detector and three-detector case,
and found excellent agreement with Monte Carlo simulations. 

The important resulting consequences are: the precision on the
background estimation saturates as a function of the number of time
slides, this saturation is most relevant for detectors with a very
different trigger rate where the background estimation precision can be poor
for any number of time slides.

Let us note that the time slide method can be used in other situations
than GW data background estimation. For example it can be used to
estimate the rate of accidental coincidences between a GW channel and
an environmental channel in a GW interferometer; or in any experiment
where coincidences between two (or more) trigger generators are looked
for. The results of this paper can be straightforwardly extended to
such an experiment.

Another limitation, the non stationarity of the data, has not been
investigated in this paper. Data non stationarity is a well known
issue in GW data analysis \cite{Mohanty00}. In the context of the time
slides method it raises the question whether the time shifted data are
still representative of the zero lag data, when large time shifts are
used. It involves both the problem of the measure of the level of data
non stationarity, and the estimation of the error it induces on the
background estimation. Further work on this issue will be the subject
of a future paper.

\appendix

\section{Two-detector integral}\label{twoITFcalc}
To compute the integral 
\begin{multline}
 \mathrm{M}_2  = \int_{x_1} \dotsi \int_{x_N} \int_{y_1} \dotsi \int_{y_N} \left[ \frac{1}{R}\sum_\pi \sum_k\cchi(x_k<p) \cchi(y_{\pi(k)}<q)  \right]^2 \\
  = \int \dotsi \int \frac{1}{R^2} \sum_{\pi_1} \sum_{\pi_2} \sum_{k} \sum_{l} 
\cchi(x_{k}<p) \cchi(x_{l}<p) \cchi(y_{\pi_1(k)}<q) \cchi(y_{\pi_2(l)}<q),
\end{multline}
we put the sums outside the integrals. When $k\neq l$, the integrals
on $x_k$ and $x_l$ are independent, and the integration over $x_1,
\dotsc, x_N $ gives a $p^2$ contribution. Otherwise the integration
gives a $p$ contribution. Analogously for the $y$ variables we get
$q^2$ or $q$ depending on whether $\pi_1(k) \neq \pi_2(l)$ or not.

Thus we get four types of integrals
\begin{subequations}
\begin{align}
 & &\qquad \text{integral} &\times \text{number of such integrals} \notag \\
 k=l,&\pi_2^{-1} \circ \pi_1 (k) = l &\qquad \frac{1}{R^2} p q &\times N R \\
 k\neq l,&\pi_2^{-1} \circ \pi_1 (k) = l &\qquad \frac{1}{R^2} p^2 q &\times N R (R -1)\\
 k=l,&\pi_2^{-1} \circ \pi_1 (k) \neq l &\qquad \frac{1}{R^2} p q^2 &\times N R (R -1)\\
 k\neq l,&\pi_2^{-1} \circ \pi_1 (k) \neq l &\qquad \frac{1}{R^2} p^2 q^2 &\times N \left[ R(R-1)(N-2) + R(N-1)\right]
\end{align}
\end{subequations}
Here we used that the composition of two circular permutation is a circular permutation, and that the only circular permutation with a fixed point is the identity. 

The details of the combinatorics are as follows. 
\begin{itemize}
 \item $ k=l,\pi_2^{-1} \circ \pi_1 (k) = l$ : There are $N$ different $k$ values. For each of them there is only one $l$ that is equal to it. Here $\pi_2^{-1} \circ \pi_1$ is a circular permutation with a fixed point, so it is the identity. There are $R$ different $\pi_1$, and for each of them only $\pi_2=\pi_1$ gives the identity. 
 \item $k\neq l,\pi_2^{-1} \circ \pi_1 (k) =l$ : There are $N$ different $k$ values. For every pair $\pi_1 \neq \pi_2$ we get $\pi_1^{-1} \circ \pi_2(k) \neq k$. And the choice of this pair determines uniquely an $l$ that is not equal to $k$. There are $R(R-1)$ such pairs.
\item $k=l,\pi_2^{-1} \circ \pi_1 (k) \neq l$ : There are $N$ different $k$ values. The value of $l$ is determined by the equality $k=l$. And there are $R(R-1)$ pairs of $\pi_1,\pi_2$ such that $\pi_1^{-1} \circ \pi_2(k) \neq k$. 
\item $ k\neq l,\pi_2^{-1} \circ \pi_1 (k) \neq l$ : There are $N$ different $k$ values. In the case where $\pi_1 \neq \pi_2$, we need that $l\neq k$ and $l \neq \pi_2^{-1} \circ \pi_1 (k)$, there are $N-2$ such $l$. In the case where $\pi_1 = \pi_2$ we get $k=\pi_2^{-1} \circ \pi_1 (k)$, so there is only one inequality on $l$, and there are $N-1$ possible $l$.
\end{itemize}

 By summing the 4 terms above and subtracting  $\mathrm{Mean}^2$ we obtain
\begin{align}
 \mathrm{Var } & = \frac{1}{R} N p q \left[ 1 + p(R-1) + q(R-1) + pq\bigl((R-1)(N-2) + (N-1)\bigr)\right] - (N p q)^2\\
& = N p q \left[ \frac{1}{R} + p + q + \frac{p q - (p + q)}{R} - 2 p q\right] \\
& \simeq N p q \left[ \frac{1}{R} + p + q\right],
\end{align}

\section{Three-detector integral}\label{threeITFcalc}
We want  to compute the integral 
\begin{multline}
 \op{M}_2 = \int \dotsi \int \frac{1}{R^2} \sum_{\pi_1} \sum_{\pi_2} \sum_k \sum_l 
\\\left[ X Y X' Y' + X Z X' Z' + Y Z Y' Z' + 4XYZX'Y'Z' + 2XYX'Z' + 2XYY'Z' \right. \\ \left. + 2XZY'Z' -4 XYX'Y'Z' - 4XZX'Y'Z' - 4YZX'Y'Z'\right],
\end{multline}
where the $'$ denotes whether the hidden variables are $\pi_1,\,k$ or $\pi_2,\,l$. 

Similarly to \ref{twoITFcalc} we have here eight kind of integrals.
\begin{equation}
\begin{matrix}
 X & Y & Z &  \text{number of such integrals} \notag \\
 k=l, & \pi_2^{-1} \circ \pi_1 (k)=l ,&\pi_2 \circ\pi_1^{-1}(k) = l,&  N R  \\
k=l, & \pi_2^{-1} \circ \pi_1 (k)=l ,&\pi_2 \circ\pi_1^{-1}(k) \neq l,&   0 \\
k=l, & \pi_2^{-1} \circ \pi_1 (k) \neq l ,&\pi_2 \circ\pi_1^{-1}(k) = l,&  0 \\
k=l, & \pi_2^{-1} \circ \pi_1 (k) \neq l ,&\pi_2 \circ\pi_1^{-1}(k) \neq l,&   N R (R-1) \\
k\neq l, & \pi_2^{-1} \circ \pi_1 (k)=l,&\pi_2 \circ\pi_1^{-1}(k) = l,&   0 \\
k\neq l, & \pi_2^{-1} \circ \pi_1 (k)=l,&\pi_2 \circ\pi_1^{-1}(k) \neq l,&   N R (R-1)\\
k\neq l, & \pi_2^{-1} \circ \pi_1 (k) \neq l,&\pi_2 \circ\pi_1^{-1}(k) = l,&   N R (R-1)\\
k\neq l, & \pi_2^{-1} \circ \pi_1 (k) \neq l,&\pi_2 \circ\pi_1^{-1}(k) \neq l,&   N R \left[(R-1)(N-3) + (N-1)\right]
\end{matrix}
\end{equation}
In these combinatoric computations we need to assume that all translations are smaller than $N/4$, to ensure that $\pi_2^{-1} \circ \pi_1 \circ \pi_2^{-1} \circ \pi_1 (k) = k \Rightarrow \pi_1 = \pi_2$. This assumption is really reasonable, and the result would not be significantly different without it.

The final result is
\begin{multline}
 \op{M}_2 =\frac{N}{R}\Bigl\{ (pq + pr + qr- 2pqr) \Bigr. \\ \left.+ (R-1)\left[ pq(p+q+pq) + pr(p+r+pr) +qr(q+r+qr) + 6pqr -4pqr(p+q+r)\right]  \right. \\ \Bigl.+ \left[ (R-1)(N-3)+(N-1) \right](pq+pr+qr-2pqr)^2\Bigr\},
\end{multline}

\section{``OR'' case for $D$ detectors}
\label{sec:orDdim}
Using the same heuristic as in section \ref{sec:2det:interpretation}
and \ref{sec:3det:interpretation} we compute the variance of the time
slide estimation method for $D$ detectors in the ``OR'' case. This
heuristic yielded the same results as the exact computation for the 2
and 3 detector case, thus we may expect it to stay true in the general case.
 
As in equation \eqref{varNb3}, the variance is
the sum of the normal Poisson variance
\begin{equation}
  \op{Var}_{\text{Poiss}} = N \left( \sum_{j=1}^D \sum_{i=1}^{j-1} p_i p_j \right) \frac{1}{R},
\end{equation}
and the variance due to time slides.

The estimate of the mean rate is 
\begin{align}
  \op{Mean} &= N \left[ \sum_{j=1}^D \sum_{i=1}^{j-1} \left( p_i + \delta p_i\right)
    \left( p_j + \delta p_j \right) \right] \\
  &\simeq N\left[ \sum_{j=1}^D \sum_{i=1}^{j-1} p_i p_j +  
  \sum_{j=1}^{D} \delta p_j \left( \sum_{\substack{i=1\\i \neq j}}^D  p_i\right) \right],
\end{align}
which leads to a variance due to multiple reuse of the data (assuming
$\langle \delta p_i^2\rangle = \frac{p_i}{N}$)
\begin{align}
  \op{Var}_{\text{Slides}}/N = &
  \sum_{j=1}^{D} p_j \left( \sum_{\substack{i=1\\i \neq j}}^D  p_i\right) 
  \left(\sum_{\substack{k=1\\k \neq j}}^D  p_k\right)\\
  = & \left( \sum_{j=1}^D \sum_{\substack{i=1 \\i \neq j}}^D p_i p_j \right) \left( \sum_{k=1}^{D} p_k \right) - 
  \sum_{j=1}^{D} p_j^2 \sum_{\substack{i=1 \\i\neq j}}^D p_i \\
  = & \left( \sum_{j=1}^D\sum_{i=1}^{j-1} p_i p_j \right) \left(\sum_{k=1}^D p_k \right) +
  \frac{1}{2} \sum_{j=1}^D\sum_{\substack{i=1\\i \neq j}}^D p_i p_j 
  \left(p_j + p_i + \sum_{\substack{k=1\\k\neq i,\, k\neq j}}^D p_k  \right) \notag \\
  &-   \sum_{j=1}^{D}  \sum_{\substack{i=1\\ i\neq j}}^D p_i p_j^2  \\
  = & \left( \sum_{j=1}^D\sum_{i=1}^{j-1} p_i p_j \right) \left(\sum_{k=1}^D p_k \right) + 
  \frac{1}{2} \sum_{j=1}^D \sum_{\substack{i=1 \\ i \neq j}}^D
  \sum_{\substack{k=1\\k \neq i,\,  k \neq j}}^D p_i p_j p_k.\label{eq:DdetORvar}
\end{align}
This general formula \eqref{eq:DdetORvar} is correctly giving back the
extra terms in equations \eqref{varNb} and \eqref{varNb3} for
respectively the 2 and 3 detector case.

\section*{References}

\bibliographystyle{unsrt}
% \bibliography{References}

\end{document}